# Advances in imaging techniques for the study of individual bacteria and their pathophysiology

Dohyeon Lee[1,2,+], Hyun-Seung Lee[4,+], Moosung Lee[1,2], Minhee Kang[5,6], Geon Kim[1,2], Tae Yeul Kim[7,*], Nam Yong Lee[7], and YongKeun Park[1,2,3,*]

[1]Department of Physics, Korea Advanced Institute of Science and Technology, Daejeon, 34141, Republic of Korea

[2]KAIST Institute for Health Science and Technology, KAIST, Daejeon, 34141, Republic of Korea

[3]Tomocube Inc., Daejeon, 34109, Republic of Korea

[4]Department of Laboratory Medicine, Wonkwang University Hospital, Iksan, 54538, Republic of Korea

[5]Biomedical Engineering Research Center, Smart Healthcare Research Institute, Samsung Medical Center, Seoul, 06351, Republic of Korea

[6]Department of Medical Device Management and Research, Samsung Advanced Institute for Health Sciences & Technology, Sungkyunkwan University, Seoul, 06351, Republic of Korea

[7]Department of Laboratory Medicine and Genetics, Samsung Medical Center, Sungkyunkwan University School of Medicine, Seoul, 06351, Republic of Korea

+Equally contributed.

*Correspondence: yk.park@kaist.ac.kr and voltaire0925@gmail.com

## Abstract

Bacterial heterogeneity is pivotal for adaptation to diverse environments, posing significant challenges in microbial diagnostics and therapeutic interventions. Recent advancements in high-resolution optical microscopy have revolutionized our ability to observe and characterize individual bacteria, offering unprecedented insights into their metabolic states and behaviors at the single-cell level. This review discusses the transformative impact of various high-resolution imaging techniques, including fluorescence and label-free imaging, which have enhanced our understanding of bacterial pathophysiology. These methods provide detailed visualizations that are crucial for developing targeted treatments and improving clinical diagnostics. We highlight the integration of these imaging techniques with computational tools, which has facilitated rapid, accurate pathogen identification and real-time monitoring of bacterial responses to treatments. The ongoing development of these optical imaging technologies promises to significantly advance our understanding of microbiology and to catalyze the translation of these insights into practical healthcare solutions.

# 1. Introduction

Bacteria are among the simplest unicellular organisms, yet they exhibit significant heterogeneity in their physiology and behavior. This heterogeneity, arising from genetic, environmental, or stochastic factors, critically impacts how bacteria adapt to and interact with their surroundings[1,2]. Understanding bacterial heterogeneity is therefore a foundational challenge in microbiology, as it plays a crucial role in bacterial survival and adaptability across diverse environments.

To investigate bacterial heterogeneity, optical microscopy techniques have been widely employed. Recently, breakthroughs in high-resolution optical microscopy have made it possible to study individual bacteria with unparalleled specificity and spatial resolution, enabling researchers to observe their metabolic states, structural organization, and dynamic interactions. These imaging technologies are transforming clinical microbiology and infectious disease research, offering rapid and precise tools for bacterial identification and real-time monitoring of bacterial responses to treatments. For example, various high-resolution microscopic techniques have been used to study the division of bacteria in biofilms, where bacteria divide different tasks to optimize their survival, such as nutrient acquisition, defense, and attachment[3-7]. Such insights can lead to new strategies for managing biofilms, a major cause of medical device-related infections.

Identifying bacterial species and determining antimicrobial susceptibility is essential for guiding treatment decisions in bacterial infections, yet traditional identification methods, including biochemical assays[8-11], serologic tests[12], staining[13,14], and conventional antimicrobial susceptibility testing (AST) methods, including broth microdilution[15], often lack speed, sensitivity, or specificity. Chromogenic cultures[16,17] and optical sensor-based approaches[8,10,18-21] offer simplicity and cost-effectiveness but require large sample volumes and may yield inaccurate AST results in mixed infections. Mass spectrometry, while rapid and affordable for many applications[20,22], still provides limited identification data due to restricted libraries[23]. Similarly, genetic methods offer highly sensitive bacterial detection with minimal sample requirements but at a high cost and with potential for therapy-related false positives[24-30]. Imaging-based technology detecting small colonies has been a potential solution for rapid bacterial identification[31,32]. These microbiological methods have been clinically tested and commercialized (Table 1). Recently, high-resolution, single-cell-based imaging has become a promising alternative because identification and AST are determined within a few cell cycles.

Moreover, high-resolution microscopy enables exploration of bacterial heterogeneity beyond simple population-level measurements, such as growth rate and biochemical composition[33-36]. By allowing single-cell visualization, these imaging methods reveal the diversity of bacterial behaviors within clonal populations and interactions between individual bacteria and external stressors. For example, high-resolution imaging has shown how bacteria can divide tasks within biofilms, enhancing their collective survival under challenging conditions.

This review discusses the transformative role of high-resolution optical microscopy in revealing bacterial properties at the single-cell level and advancing our understanding of bacterial functions. We begin with an overview of high-resolution microscopic techniques, from fluorescence to label-free imaging, and explore their contributions to real-time bacterial identification, locomotion analysis, biofilm research, and monitoring cellular responses to environmental changes. Additionally, we highlight the potential for integrating these imaging techniques with microfluidic platforms and advanced image processing, which promises to further enhance their impact on both preclinical research and clinical diagnostics.

**Table 1 | Microbiological techniques for bacterial identification and AST for bloodstream infections in clinical microbiology laboratories**

| Mechanism | Method or product | Manufacturer | Automation | Samples | | | Turnaround time | AST | Advantage | disadvantage |
|---|---|---|---|---|---|---|---|---|---|---|
| | | | | Blood | Culture-positive broth | Sub-culture | | | | |
| Staining | Manual staining[13,14] | - | X | . | O | O | 0.2 h | X | Relatively rapid and simple. Cost-effective. | Labor intensive. Low sensitivity/specificity. Needs confirmatory tests. |
| Biochemical test | Manual biochemical test | - | X | . | . | O | | X | Cost-effective. | Time-consuming, labor-intensive. Complicated procedures. Requires large numbers of bacteria. |
| | API[8-10] | bioMerieux | X | . | . | O | 4–24 h | X | | |
| | BBL Crystal[9] | BD Diagnostics | X | . | . | O | 4-20 h | X | | |
| | RapID[11] | Thermo Fisher | X | . | . | O | 4-6 h | X | | |
| Serologic test | Manual serologic test[12] | - | X | . | O | O | 1-3 day | X | Alternative when pathogen culture is difficult. | Time-consuming, Requires acute and convalescent specimens. |
| Chromogenic culture | CHROMID[16] | bioMerieux | X | . | O | O | 24 h | O | Simple, standardized method. | Time-consuming. Requires additional confirmatory tests. |
| | Spectra MRSA[17] | Thermo Fisher | X | . | O | O | 24 h | O | | |
| Optic sensor-based | MicroScan WalkAway[8,10,18,19] | Beckman Coulter | O | . | . | O | 6–16 h | O | Cost-effective. Detects resistance mechanisms. | Requires large numbers of bacteria. Discrepancy in AST results. Different test kits for each type of bacteria. |
| | VITEK[10,18-20] | bioMerieux | O | . | . | O | 6–16 h | O | | |
| | Sensititre ARIS[21] | Thermo Fisher | O | . | . | O | 18–24 h | O | | |
| | BD Phoenix[10,19] | BD Diagnostics | O | . | . | O | 6–16 h | O | | |
| Mass spectrometry | VITEK MS[20,22] | bioMerieux | X | . | O | O | <0.5 h | X | Accurate and rapid. Cost-effective. | Requires large numbers of bacteria. Additional procedure for bacteria concentration, when culture positive broth was directly used. Inaccuracy in combined infections. Requires an additional procedure for AST. |
| | Biotyper[22] | Bruker Daltonics | X | . | O | O | <0.5 h | O | | |
| Genetic test | T2MR[24] | T2 Biosystems | O | O | . | . | 3–5 h | X | Requires small numbers of bacteria. Accurate and rapid. AST results with resistance mechanism detection. Panel flexibility with multiplex PCR and primer design modification. | High cost. Various sensitivities according to structure of certain bacterial matrix. Therapy-related false positive. Inaccuracy in combined infection. Discrepancy of AST results compared to the reference method. |
| | Magicplex Sepsis real-time test[25] | Seegene | X | O | . | . | 6 h | O | | |
| | iDTECT Dx Blood[26] | PathoQuest | X | O | . | . | 2–3 days | O | | |
| | LightCycler SeptiFast[27] | Roche | O | O | . | . | 3–4 h | X | | |
| | SepsiTest[28] | Molzym | O | O | . | . | 8–12 h | X | | |
| | FilmArray BCID2[29] | bioMerieux | O | . | O | . | 1 h | O | | |
| | GeneXpert MRSA/SA Blood Culture[30] | Cepheid | O | . | O | . | 1 h | O | | |
| Imaging | Accelerated Pheno system[31] | Accelerate Diagnostics | O | . | O | . | 7 h | O | Requires small numbers of bacteria. Rapid. AST result with resistance mechanism detection | Requires fresh culture-positive broth. Discrepancy in AST results. |
| | dRAST[32] | QuantaMatrix | O | . | O | . | 5–7 h | O | | |

## 2. Recent advances in high-resolution microscopy for live bacterial imaging

The transparent nature of unlabeled bacterial cells presents a considerable challenge for researchers aiming to achieve high-contrast imaging. Traditional approaches, such as staining and fluorescence labeling, have been employed to visualize bacterial cells and their substructures with enhanced contrast or molecular specificity. However, fixed-cell imaging methods are inherently limited, as they can only capture dead cells. Additionally, genetic transfections for fluorescence imaging may alter the physiological characteristics of live bacteria and introduce issues such as photobleaching and phototoxicity.

To address these limitations, various non-invasive optical microscopy techniques have been developed, allowing researchers to capture the dynamic activities of live bacterial cells. Nonetheless, achieving high spatiotemporal resolution and sufficient contrast to visualize substructures and molecular details—particularly in motile bacteria—remains a significant challenge.

Recent advances in microscopy have successfully overcome many of these obstacles, enabling high-resolution imaging of living bacterial cells. These techniques can be broadly categorized into two classes: high-resolution fluorescence microscopy and label-free microscopy. High-resolution fluorescence microscopy requires labeling, while label-free techniques do not rely on any exogenous markers. Pushing the technical boundaries of both spatial and temporal resolution has opened new possibilities for in-depth studies of subcellular structures, bacteria-bacteria interactions, and long-term investigations of bacterial pathophysiology.

In this section, we will explore recent advances in high-resolution microscopy for live bacterial imaging (Fig. 1). We will discuss how these innovations have enabled researchers to surpass the challenges of imaging unlabeled bacterial cells with high spatiotemporal resolution. This includes developments in super-resolution microscopy, structured illumination microscopy, and light-sheet microscopy. Furthermore, we will cover label-free techniques, such as phase-contrast microscopy, differential interference contrast microscopy, and quantitative phase imaging, which facilitate the imaging of live bacteria without the need for labeling.

### 2.1 High-resolution fluorescence microscopy

Fluorescence microscopy is a well-established tool in biological research, allowing for the visualization of specific organelles and molecules by attaching different fluorescent markers to multiple objects of interest. This enables the observation of separate structures through multi-channel imaging. Alongside fluorescence microscopy, non-invasive optical microscopy techniques have been developed to visualize color-tagged bacterial cells while keeping them alive. The primary goal of high-resolution fluorescence microscopy for live bacterial imaging has been to enhance spatiotemporal resolution and extend the imaging window, effectively increasing the spacetime-bandwidth product. Techniques such as optical sectioning microscopy, non-uniform excitation microscopy, and single-molecule microscopy exemplify the advancements made in these areas.

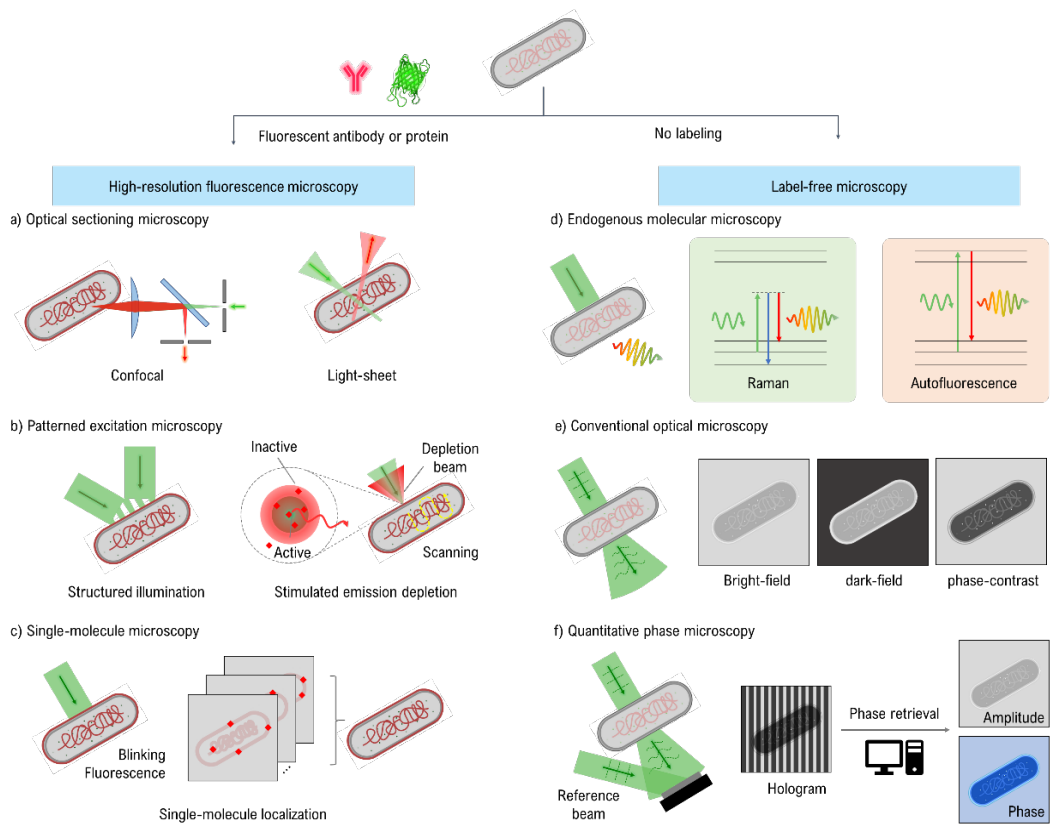

**Fig. 1** Schematic diagrams of high-resolution microscopy techniques are categorized based on their dependence on bacterial sample labeling. (a-c) High-resolution fluorescence imaging techniques require sample labeling before microscopic imaging. (a) Optical sectioning microscopy removes fluorescent signals outside the focal plane for high-contrast imaging. (b) Nonuniform excitation microscopy achieves high-resolution imaging utilizing an engineered excitation beam. (c) Single-molecule microscopy tracks the exact positions of blinking fluorophores in a few nanometer scales. (d-f) Label-free imaging techniques enable the imaging of raw bacterial samples. (d) Endogenous molecular microscopy utilizes the excitation-relaxation of specific endogenous molecules to generate optical fingerprints. (e) Conventional optical microscopy and (f) quantitative phase imaging (QPI) visualize bacteria by contrasting or visualizing light-scattering patterns produced by the bacterial cells.

*Optical sectioning microscopy*

Optical sectioning microscopy enhances the contrast of images by suppressing fluorescence noise outside the focal plane (Fig. 1a). Confocal microscopy and light sheet microscopy are two popular techniques in high-resolution fluorescence microscopy for imaging live bacterial cells. Confocal microscopy uses a laser beam to excite fluorescently labeled molecules in a sample, and a pinhole in front of the detector to selectively capture the emitted light from the focal plane[37]. This technique provides high contrast and spatial resolution by removing out-of-focus light, which results in a clearer image of the sample. However, confocal microscopy is relatively slow and may cause phototoxicity and photobleaching due to its high-power laser excitation.

Light sheet microscopy, on the other hand, uses a thin sheet of light to excite the fluorescently labeled sample in a plane perpendicular to the direction of imaging[38]. The emitted light is then detected by a camera positioned perpendicular to the plane of illumination, which allows for fast and gentle imaging of living bacterial cells. This technique has the advantage of high-speed imaging, high contrast, and minimal photodamage, making it an excellent option for long-term imaging of live cells. Light sheet microscopy is particularly useful for studying bacterial dynamics, such as in biofilms, where it can provide a 3D view of the sample.

Lattice light sheet microscopy is an advanced version of light sheet microscopy that employs a lattice pattern of light sheets for even greater imaging speed and reduced phototoxicity[39]. Lattice light sheet microscopy allows for ultrafast imaging at up to 1,000 frames per second, which is fast enough to capture rapid cellular processes such as bacterial motility and cell division. Additionally, the use of multiple light sheets in lattice light sheet microscopy results in less photodamage to the sample since each individual sheet is lower in intensity than a single sheet in traditional light sheet microscopy. This makes it possible to image live cells for longer periods without damaging them.

*Nonuniform excitation microscopy*

Nonuniform excitation microscopy yields super-resolution by giving gated illumination onto the sample through spatially nonuniform light (Fig. 1b). Structured illumination microscopy (SIM) uses sinusoidal illuminations to evoke the mixing of frequencies between harmonic patterns and the sample. Reconstruction of the digital image reveals previously inaccessible high-frequency components encoded into the observed image, providing lateral resolution beyond the diffraction limit. To further improve the resolution, saturated (nonlinear) SIM can be employed by illuminating high-powered sinusoidal beams and saturating the fluorescent response[40,41]. However, phototoxicity and photobleaching problems limit its applications. Alternatively, stimulated emission depletion (STED) microscopy is one of the earliest super-resolution methods widely utilized in microbiology until now[42].

In contrast with nonlinear SIM, STED microscopy sends the fluorophores into a dark state except for the focal spot. This de-excitation or depletion of the surrounding emission can be achieved using a high-powered donut-shaped depletion beam. Because it is a scanning-based method, STED microscopy takes advantage of rapid super-resolution imaging of bacteria by restricting the field of view. Additionally, photostable[43] or reversely photo-switchable[44] proteins can reduce photobleaching and phototoxicity problems for live-cell imaging.

*Single-molecule microscopy*

Single-molecule localization microscopy (SMLM) is another super-resolution method that exploits the stochastic emission process of fluorophores (Fig. 1c). By separating blinking fluorophores from time-lapse fluorescence images, SMLM enables super-resolution without specialized instrumentation[45,46]. Rapid-acquisition SMLM enables visualization of spatially clustered protein distributions, chromosome dynamics tracking, and protein molecule diffusion in living bacteria[47-51].

## 2.2. Label-free microscopy

Photobleaching and phototoxicity are the significant hurdles of fluorescence microscopy that prevent its widespread use in microbiology. Furthermore, conventional fluorescent proteins require sufficient oxygen to emit light, thus making this technique incompatible with anaerobic bacteria[52]. Several label-free approaches have been devised to capture the intrinsic properties of living bacteria to overcome such challenges.

*Phase contrast microscopy*

Throughout the history of microscopy, scientists have relied on scattered light to visualize unlabeled bacteria (Fig. 1e). Bright-field microscopy, phase-contrast microscopy, and differential interference contrast (DIC) microscopy are widely used label-free techniques for imaging bacteria. Bright-field microscopy is a simple and widely used optical microscopy technique that provides contrast based on the absorption and scattering of light by the sample. While it is useful for observing the overall morphology of bacterial cells and providing information about bacterial growth and division, it lacks high image contrast due to weak scattering of bacteria, resulting in poor contrast of the refractive index of bacteria and their surrounding media.

Phase-contrast microscopy enhances contrast between objects of similar refractive index, enabling visualization of live bacteria without labeling[53]. This method is particularly useful for observing bacterial motility and flagellar movement, providing insights into the mechanisms of bacterial motility. Differential interference contrast (DIC) microscopy utilizes polarized light to generate contrast based on variations in refractive index and optical path length of the sample[54]. This technique enhances the contrast of edges and boundaries, making it particularly useful for observing bacterial structures and the interactions between bacteria and their environment. Label-free phase-contrast techniques allow for imaging live bacteria without fluorescent labeling or genetic modifications, providing a valuable tool for studying bacterial physiology and behavior. However, they only provide qualitative imaging information and are limited to 2D.

*Label-free molecular imaging*

Autofluorescence and Raman scattering microscopy are two label-free imaging techniques that are becoming increasingly popular in microbiology. Autofluorescence microscopy takes advantage of the intrinsic fluorescence of certain endogenous biomolecules in bacterial cells, such as NADH, flavins, and porphyrins, to generate contrast and image live bacteria without the need for exogenous labels[55]. The emitted fluorescence spectra can also provide information on the metabolic state of the bacteria[56], allowing for the monitoring of bacterial physiology and behavior over time. Via fluorescence lifetime microscopy (FLIM), auto-fluorophores could be able to unravel the internal condition of bacterial cells non-invasively[57]. Autofluorescence microscopy has been used to study a variety of bacteria, including monitoring citrus greening diseases in plants caused *by Candidatus Liberibacter asiaticus*[58], laser-induced fluorescence spectroscopy imaging of murine gastrointestinal tract[59].

Raman scattering microscopy is another label-free technique that provides information on the chemical composition of bacterial cells[60,61]. Raman scattering occurs when incident light interacts with chemical bonds in the sample, leading to a shift in the energy of the scattered photons. By analyzing the spectrum of the scattered light, the chemical composition of the sample can be determined. Raman scattering microscopy can provide molecular specificity and has been used to image bacterial cells and study their interactions with their environment. A recent study took advantage of chemical specificity through deuterium-tagging and used this to image the metabolic activity of live bacteria upon antibiotics susceptibility test[62-66]. However, the technique can be challenging due to the low signal intensity of Raman scattering and the potential for sample damage from the high-intensity laser used for excitation. Advances in Raman scattering microscopy, such as surface-enhanced Raman scattering (SERS) and coherent Raman scattering (CRS), have improved sensitivity and reduced photodamage, making it a promising tool for studying bacterial physiology and behavior[60,67].

*Quantitative phase imaging*

Quantitative phase imaging (QPI) is a label-free technique that exploits refractive index as an intrinsic quantitative imaging contrast[68] (Fig. 1f). 2D QPI techniques, such as interferometric microscopy or digital holographic microscopy, measure the optical phase delay that provides information about the optical thickness and refractive index of biological samples. 3D QPI techniques, also known as holotomography[69], reconstruct the 3D RI distribution of a biological sample from multiple 2D optical measurements.

QPI is a powerful tool for observing living bacteria due to its high sensitivity and ability to capture fine details of cells and their behavior without the need for labeling or staining. This technique has been extensively used in various applications, including morphological and structural analysis[70], and growth kinetics[49]. Quantitative measurement of the refractive index of bacteria and their surroundings by QPI allows for the precise detection of small changes in cell morphology and dynamics[71], making it a valuable tool for studying bacterial physiology and behavior. Holotomography, one of the 3D QPI techniques, has enabled the quantification of polyhydroxyalkanoates (PHA) produced by individual bacteria, which includes the concentration, volume, and dry mass of PHAs in live unlabeled bacteria[72], and the investigation of antibacterial activities of engineered films[73] or applied antibiotics[74] by monitoring the dynamics of individual bacteria.

Moreover, QPI has potential applications in clinical diagnostics. For instance, QPI has been utilized to detect anthrax spores by analyzing 2D optical field images of individual bacterial spores using a machine-learning approach[75]. High-resolution label-free imaging of the growth of individual bacteria upon antibiotic treatments by QPI also suggests potential for rapid label-free image-based antibiotic susceptibility testing (AST)[74]. Holotomography, coupled with a deep-learning algorithm, was used to identify bacterial pathogens commonly found in sepsis[76]. Hence, QPI offers a promising alternative to traditional staining and culturing methods for rapid detection and identification of bacterial pathogens in clinical settings.

*Electrical imaging*

Label-free electrical techniques represent an exciting frontier in the study of cellular systems. These techniques provide impedance and electrochemical information on cellular properties, including redox potential[77], cell–cell adhesion[78], and real-time kinetics[79], which are often challenging to capture with optical imaging methods. Due to their label-free nature, electrical imaging techniques offer the unique advantage of monitoring cellular behavior over extended periods with minimal perturbation to cellular physiology. This makes them particularly well-suited for investigating microbial populations' responses to varying environmental conditions or AST[80].

Recent advancements in complementary metal-oxide-semiconductor (CMOS) microelectrode array (MEA) technology have enabled high-resolution, high-throughput functional imaging of unlabeled live-cell cultures through *in situ* impedance and electrochemical measurements[81-83]. CMOS MEAs provide an unparalleled opportunity to capture the real-time responses of adherent cells, such as those found in biofilms, with remarkable precision. As the feature sizes of CMOS chips continue to shrink, the spatial resolution of electrical imaging is expected to further improve, allowing for subcellular characterization of bacteria. This opens new avenues for exploring microbial physiology at an unprecedented level of detail.

## 3. High-resolution imaging techniques for the study of bacterial physiology

High-resolution imaging techniques can be used to identify bacteria and study their physiological properties, such as locomotion activity, biofilm formation, and cellular response. Fluorescence-tagged molecules can provide information about subcellular structures or represent the metabolic state of the cell. Autofluorescence imaging of endogenous markers, such as NADH/NAD+, can reveal cellular metabolism through redox reactions. Quantitative phase imaging (QPI) can provide dry mass information through phase shift measurements, which have a linear relationship with protein concentration. These techniques have been actively studied in the microbiological field to gain insights into bacterial physiology and behavior. In the following subsections, we describe details of such applications that are actively studied in the microbiological field.

## 3.1 Bacterial species identification

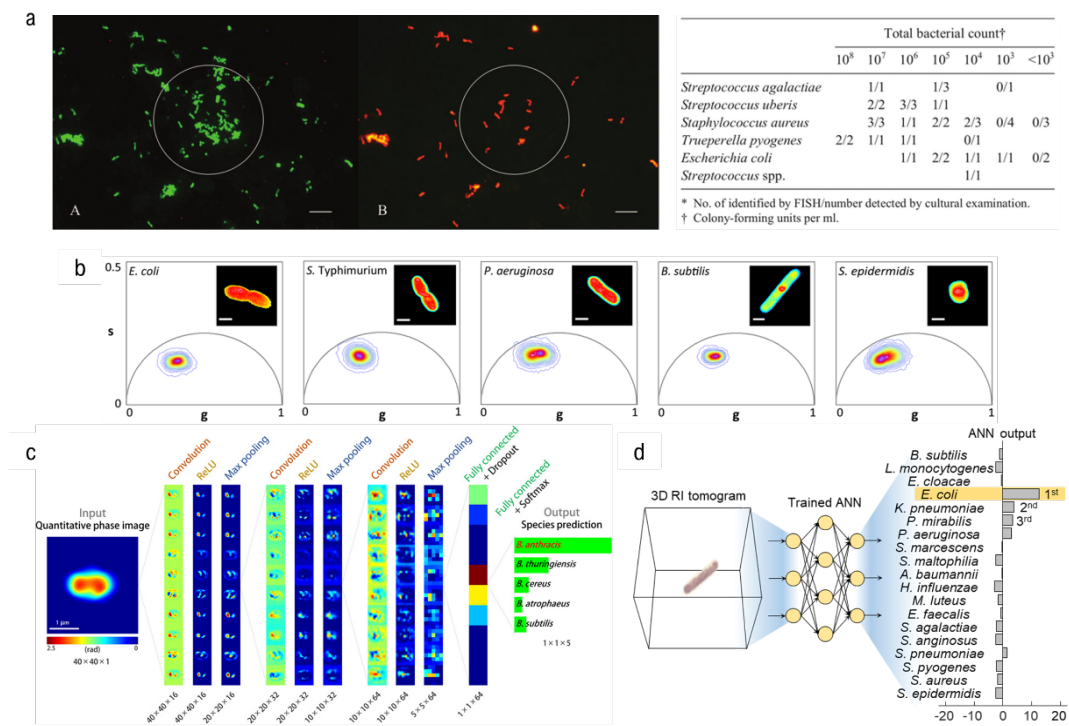

**Fig. 2** (a) FISH analysis detects bacteria in a species-specific manner through the usage of specific fluorescent probes. (b) Bacterial phasor fingerprint map generated by FLIM, where phasor shifts are considered as metabolic changes within the cells. (c) Identification of anthrax spores using custom neural network. (d) A neural activation profile resulted from artificial neural network processing of a 3D refractive index tomogram. (a) is adapted from Ref.[84]. (b) is adapted from Ref.[85]. (c) is adapted from Ref. [75]. (d) is adapted from Ref.[76].

The examination of bacteria using conventional optical microscopy cannot distinguish between different bacterial species, as they exhibit similar shapes and sizes. However, bacterial species identification is crucial for the diagnosis and treatment of infectious diseases, and various identification methods are used in clinical microbiology laboratories. Matrix-assisted laser desorption/ionization time-of-flight (MALDI-TOF) mass spectrometry has been widely used for bacterial species identification, but it requires high initial cost for MALDI-TOF equipment[86].

Fluorescence *in situ* hybridization (FISH) employed imaging has been proven to be a reliable method for bacterial detection under complex environments since fluorescence tagging enables chemical imaging with high specificity. FISH targets a specific sequence of DNA or RNA using complementary probes, providing chemical imaging with high specificity[87]. FISH has been used to detect bacteria in various applications, ranging from quantifying microbial populations in beverages[84] to laboratory diagnostics[88] (Fig. 2a). Smartphone-based FISH microscopy has also been introduced, further extending the application scope[89].

Despite the high specificity of FISH microscopy, it is limited to identifying only the bacterial species that have available probes. As an alternative, autofluorescence microscopy can be used for label-free bacterial identification. Autofluorescence is an intrinsic modality of fluorescence emitted by molecules naturally residing in cells. The potential of autofluorescence microscopy for bacterial detection was demonstrated by selectively imaging pathogenic bacteria, including Mycobacterium tuberculosis[90]. In a recent study, biomarkers specific to different bacterial species were found in the autofluorescence images obtained under a two-photon setting of fluorescence lifetime imaging microscopy (FLIM)[85] (Fig. 2b). The fingerprint patterns of five different species were located in the phasor of FLIM measurement, which resulted from the autofluorescence of metabolic molecules.

QPI imaging provides label-free images that offer direct evidence for certain bacterial species. The rich morphological profiling obtained from QPI can be statistically leveraged using machine learning-based classification. Jo et al. suggested the potential of QPI for bacterial identification using linear machine learning to classify bacterial species from angular spectrum features that were made accessible through QPI measurements[91]. Another study successfully distinguished spores of multiple *Bacillus* species, including the hazardous *B. anthracis*, by introducing deep learning to classify images acquired with a portable QPI unit[75] (Fig. 2c). Recently, bacterial identification using QPI has been advanced by combining 3D QPI and an advanced 3D deep learning architecture[76] (Fig. 2d). The high label-free single-cell identification performance was exploited to distinguish among 19 bacterial species of bloodstream infection pathogens.

## 3.2 Bacterial locomotion

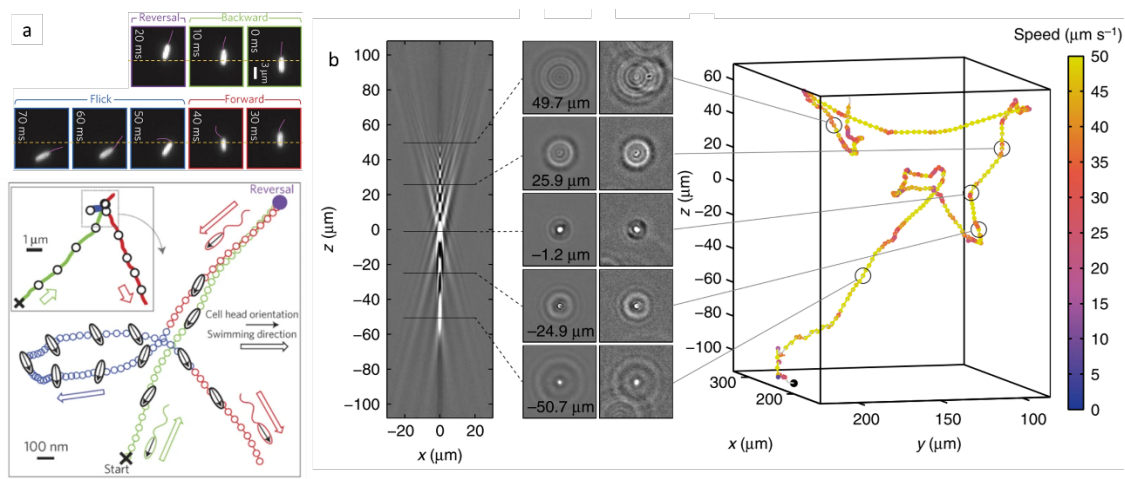

**Fig. 3** (a) Dark-field microscopy capture an image sequence of a single bacteria at 420 frames per second. Bacteria show three different motility phases. (b) Conventional phase contrast 3D tracking of a single bacterium without mechanical refocusing. They measured the axial position of bacteria via cross-correlation between diffracted patterns and a reference library. (a) is adapted from Ref.[92,93]. (b) is adapted from Ref.[94].

Measuring the locomotion of individual bacteria is important for understanding bacterial behavior, interactions, and responses to stimuli[95]. High-resolution optical microscopy has been used to study bacterial motility at the individual cell level, allowing for the measurement of the speed, direction, and trajectory of bacterial movement. These techniques have provided insights into the mechanisms of bacterial motility and the behavior of bacterial populations in response to environmental factors.

Representative studies using high-resolution optical microscopy for measuring the locomotion of individual bacteria include the use of phase contrast microscopy to study the behavior of flagellar motors[92,96,97], the use of dark-field microscopy to observe the movements of bacteria in response to chemical stimuli, and the use of high-speed fluorescence microscopy visualize flagellar motion[98,99]. To identify the bacterial motion in a real environment, several studies analyze super-diffusive properties in colloidal media[100,101] or colonies[102]. Phase contrast microscopy and QPI methods have also been utilized to extract the longitudinal position of bacteria[94,97,103-105] (Fig. 3b). These techniques have provided valuable information on the behavior of individual bacteria and their responses to various stimuli, providing insights into the complex interactions and dynamics of bacterial populations.

### 3.3 Bacterial morphology and biofilm formation

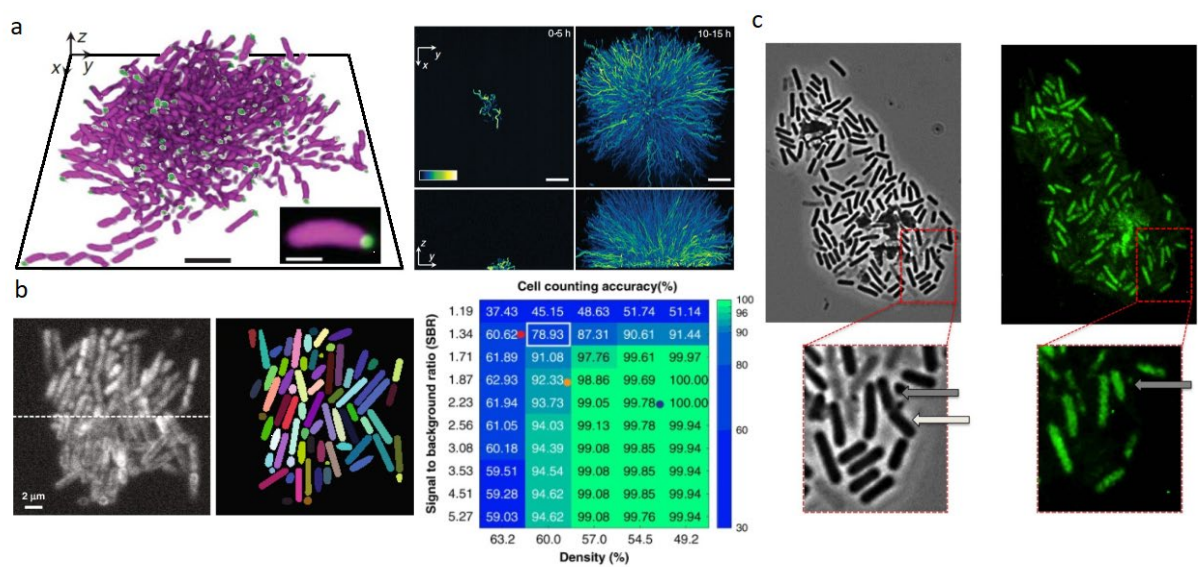

**Fig. 4** (a) Light-sheet microscopy reveals single-cell morphology under cytoplasmic punctum staining enabling tracking of each cell in colony formation. (b) Deep learning was applied for single bacterial cell morphology tracking in light-sheet microscopy. (c) Metabolic status from autofluorescence signals display the heterogenous feature in a monoclonal group. (a) is adapted from Ref.[106]. (b) is adapted from Ref.[107]. (c) is adapted from Ref.[108].

Three-dimensional (3D) single-cell imaging has revolutionized the study of bacterial morphology and behaviors. For instance, 3D structured illumination microscopy (SIM) has revealed a torus-like topological structure of the circular chromosome inside living Escherichia coli, known as heterogeneous[109].

Biofilms, the macroscopic multicellular communities formed by bacteria, represent another critical area of microbiological research. The aggregate of bacteria provides shelter from harsh environments or agents, and the matrix is optimal for capturing nutrients for their growth[6]. The safe microbial shelter challenge to solve biofouling and biocorrosion issues[4] or bacterial disease[110]. Deciphering the successful cooperative behavior of bacteria may also provide effective countermeasures against bacteria. Biofilms have been studied using volumetric fluorescence microscopy techniques that have recently circumvented the limitations of conventional 2D fluorescence microscopy.

Fluorescence microscopy is a prevalent tool for biofilm imaging, allowing optical sectioning microscopy with 3D resolution to reduce noise from neighboring bacteria outside the focal plane. The technological advances enabled quantitative analysis of 3D biofilm dynamics[111]. Spinning disk confocal microscopy has been used to track thousands of cells[112] and the disrupted collectivity of biofilms upon antibiotic treatment[113]. Similarly, light-sheet fluorescence microscopy has revealed the fountain-like flow of bacterial communities during biofilm formation[106] (Fig. 4a). Cell counting and segmentation of light-sheet fluorescence images of living biofilms can be effectively carried out using a deep learning-based image analysis workflow[107] (Fig. 4b). Autofluorescence studies have identified the metabolic activities of individual bacteria while forming biofilms[108], providing insight into the mechanism of biofilm development (Fig. 4c). Bimodal pigmentation capacity in *Bacillus pumilus* SF214 has been found to exhibit two types of cells, each group contributing differently to biofilm formation. Studying cooperative dynamics at a single-cell resolution will provide better understanding of the mechanisms of biofilm development[114,115].

## 3.4 Monitoring cellular dynamics

Modifying the genetic composition and expression profiles of bacterial species to monitor their metabolic pathways is of great interest in microbiology. Bacteria can convert carbon and nitrogen sources into various intracellular and extracellular biopolymers, so regulating or designing this cell factory can have benefits such as pathogenesis modulation and material production[116]. PHA biopolymer accumulation processes were compared in two different recombinant species[117].

## 3.5 Image-based AST

AST is routinely performed in clinical microbiology laboratories to guide selection of appropriate antibiotics. Conventional AST methods including the disk diffusion test, Epsilometer test, and broth dilution test are simple, but they require considerable time for growth feature inspection under the presence of antibiotics. High-resolution imaging tools, on the other hand, can enable the observation of phenotypic changes of bacterial cells at a single-cell level within a few doubling times. By monitoring bacterial growth from the very beginning, the culturing time needed to obtain AST results can be reduced.

Various microscopic imaging approaches have been used to provide rapid AST results. Fluorescence microscopy is an imaging technique widely used for rapid AST. Lu et al. demonstrated AST at a single-cell level by analyzing bacterial growth using time-lapse fluorescence images of individual bacteria loaded into microchannels under various antibiotic environments[84,118] (Fig. 5a). Mohan et al. proposed a similar approach to develop a microfluidic biosensor platform[119]. They fabricated a microfluidic chip in the form of an array in which both the antibiotics and the concentration were varied, and fluorescence images were used to monitor cell growth and death.

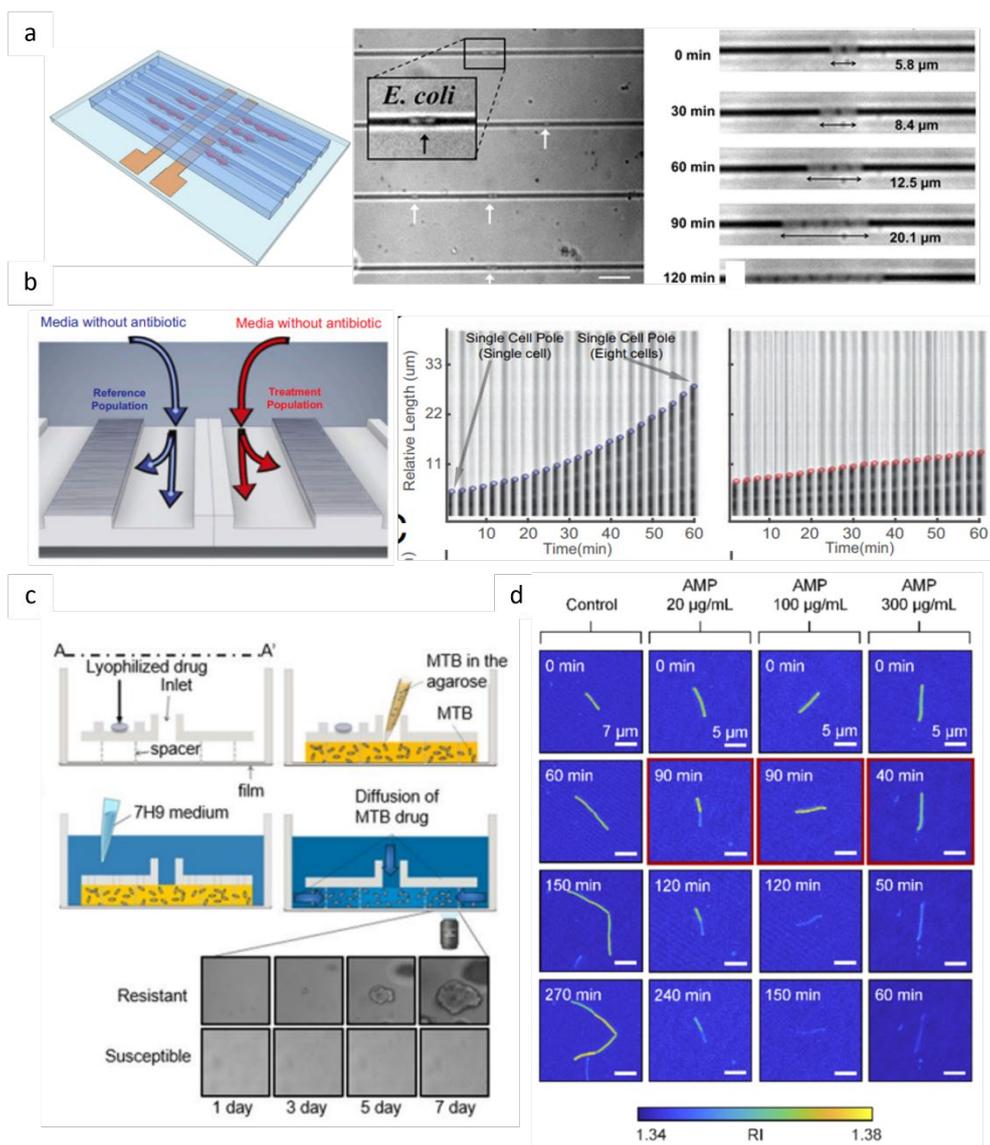

**Fig. 5** High-resolution imaging techniques used for antimicrobial susceptibility testing (AST). (a) Single-cell AST performed on E. coli in microchannels provides rapid determination of antimicrobial-resistant profiles. (b) Growth rate comparison between single-cell traps with and without antibiotics enables fast AST in less than 30 minutes. (c) The disc agarose channel system is validated for drug susceptibility testing of M. tuberculosis strains. (d) Time-lapse 3D imaging through optical diffraction tomography reveals the response of B. subtilis to ampicillin treatment. (a) is adapted from Ref.[84,118]. (b) is adapted from Ref.[120]. (c) is adapted from Ref.[121,122]. (d) is adapted from Ref.[74].

Recently, Label-free imaging methods, including bright-field microscopy, have also been used to assess antibiotic susceptibility in conjunction with similar micro-loading platforms. Matsumoto et al. assessed drug susceptibility using bright-field images from a microfluidic agarose channel chip, with bacterial number and size used to carry out AST in agreement with the broth dilution test[123]. Another AST method based on bright-field microscopy utilized arrays of microscopic wells to efficiently and multiplexedly monitor bacteria in various antibiotic environments. Veses-Garcia et al. seeded and monitored bacteria in over 600 conditions using a slide of nanoliter-sized wells[124]. Baltekin et al. devised a microfluidic trap to monitor bacterial growth rate[120] (Fig. 5b), while Jung et al. proposed micro-culture channels to determine critical drug concentrations for *Mycobacterium* *tuberculosis*[121,125] (Fig. 5c). Alternatively, QPI can also characterize bacteria under antibiotic influence, with time-lapse analysis of 3D QPI measurements indicating the effectiveness and dose of antibiotics[36,74] (Fig. 4d). More recently, a blood culture-free ultra-rapid AST platform was reported[126], which bypasses traditional blood culturing and performs susceptibility profiling directly from whole blood. This approach reduced the turnaround time by more than 40–60 hours compared to conventional AST workflows, demonstrating significant potential for faster clinical decision-making in critical cases such as sepsis.

## 4. Discussion

We explored the latest innovations in microscopy techniques and invited scientists from diverse research areas to bring novel bioengineering tools to solve new questions in microbiology. Specialized fluorescence tagging and label-free approaches provide endogenous optical fingerprints that allow us to observe biomolecular processes and metabolic states of individual bacteria, respectively. High-resolution imaging methods enable the real-time visualization of bacteria trajectories in 2D and 3D space, such as monitoring the construction process of biofilms, where the same types of bacteria participate in constructing their habitat by taking on specialized tasks. Understanding the mechanism and importance of functional specialization will unveil the strategy of bacteria to survive in a harsh environment. In the clinical approach, the advance and commercialization of culture-free identification and monitoring systems will significantly reduce the death rates due to pathogenic diseases.

Compared to other methods, such as molecular or mass spectrometric approaches, the novel image-based optical techniques have shown competitive performance for rapidity and accuracy of bacterial identification and AST[126,127]. Significant efforts have been made to improve the performance of these techniques to the level comparable to that of the reference methods of bacterial identification and AST; however, several clinical studies reported limitations of these techniques for identification in multi-bacterial infections and AST for specific bacteria, such as *Pseudomonas aeruginosa*[128]. Further improvements and research might be required to meet clinical needs.

A microfluidic approach will ultimately be a key technique for assessing and sorting a multitude of bacteria for automated monitoring. While high-resolution microscopy provides detailed morphological traits, the small field of view limits the throughput of the setups. A microfluidic chip platform scales down the monitoring system into the microscope window, which relieves the throughput issues. The miniature processing unit tests the small size of samples efficiently. The potential progress of antibiotic susceptibility tests by using a micro-loading chip has been described[118,119,123], which enables the automatic analysis of morphological changes in single bacteria under various antimicrobial conditions[129]. Recently, microfluidic chips have been used to generate a gradient of antibiotic concentration to determine the minimal inhibitory concentration[130,131]. The combination of high-resolution imaging modalities with the fine fluidic controller will facilitate the development of an automated high-throughput monitoring system.

While deep learning-based microbiology has proven useful alongside various imaging modalities, label-free imaging particularly attracts the utilization of deep learning. The label-free image provides consistent data suitable for deep learning by profiling bacteria without perturbations irrelevant to physiology. This is in contrast to the label-based image contrast, which suffers from signal degradation and stochastic errors during labeling or imaging.

Consistency in data acquisition prevents overfitting issues, as the discrepancy between the training data and new data can be minimized without the above-mentioned errors. This advantage in data quality highlights label-free imaging as a favorable option in deep image-based microbiology.

One of the challenges in adopting deep learning for image analysis is the limited interpretability of its underlying working principle. It will be essential to allow human experts to understand, review, and adjust the neural networks to scale up deep image-based microbiology and satisfy the reliability required for a biomedical assay. However, accounting for the network operation remains perplexing, particularly since the designs for neural networks have become more complex for high performance over time[122,132,133]. Early heuristic methods to locate the network's region of interest[134,135] have been followed by a pioneering implementation of Bayesian deep learning that demonstrated the quantification of uncertainties lying in the data and network[136]. Additionally, researchers have started to suggest the need and candidates for a public platform to share and evaluate the data and network models[137]. We believe that the technical advances that unveil the operation of neural networks and the establishment of a universal platform will accelerate the extensive biomedical application of deep image-based microbiology.

Overall, the innovations in high-resolution microscopy techniques discussed in this review have provided new insights and tools for studying bacterial morphology, physiology, and antibiotic susceptibility. These advances have the potential to significantly impact both basic research and clinical applications. Looking ahead, there is great potential for further improvements and integration of these imaging modalities with microfluidic technologies, deep learning, and other novel approaches. As the field of microbiology continues to evolve, we believe that the continued development and application of high-resolution imaging techniques will play an increasingly important role in advancing our understanding of bacterial biology and improving human health.

*List of abbreviations*

AST: antimicrobial susceptibility testing, STED: stimulated emission depletion, SIM: structured illumination microscopy, SMLM: Single-molecule localization microscopy, FISH: Fluorescence in situ hybridization, FLIM: fluorescence lifetime microscopy, QPI: quantitative phase imaging, 3D: Three-dimensional

*Declarations*

*Availability of data and materials*

Not applicable.

*Competing interests*

The authors declare no competing interest.

*Funding*

This work was supported by National Research Foundation of Korea (RS-2024-00442348, 2015R1A3A2066550, 2022M3H4A1A02074314), and Korea Institute for Advancement of Technology (KIAT) through the International Cooperative R&D program (P0028463).

*Authors' contributions*

All authors contributed to writing subsections and revised the manuscript. NYL and YKP supervised the projects.

*Acknowledgments*

Not applicable.

*References*

1   Davis, K. M. & Isberg, R. R. Defining heterogeneity within bacterial populations via single cell approaches. *Bioessays* **38**, 782-790 (2016).
2   Dewachter, L., Fauvart, M. & Michiels, J. Bacterial heterogeneity and antibiotic survival: understanding and combatting persistence and heteroresistance. *Molecular cell* **76**, 255-267 (2019).
3   Hall-Stoodley, L., Costerton, J. W. & Stoodley, P. Bacterial biofilms: from the natural environment to infectious diseases. *Nat Rev Microbiol* **2**, 95-108 (2004). https://doi.org/10.1038/nrmicro821
4   De Carvalho, C. C. Marine biofilms: a successful microbial strategy with economic implications. *Frontiers in marine science* **5**, 126 (2018).
5   Koo, H., Allan, R. N., Howlin, R. P., Stoodley, P. & Hall-Stoodley, L. Targeting microbial biofilms: current and prospective therapeutic strategies. *Nat Rev Microbiol* **15**, 740-755 (2017). https://doi.org/10.1038/nrmicro.2017.99
6   Flemming, H. C. *et al.* Biofilms: an emergent form of bacterial life. *Nat Rev Microbiol* **14**, 563-575 (2016). https://doi.org/10.1038/nrmicro.2016.94
7   Van Acker, H., Van Dijck, P. & Coenye, T. Molecular mechanisms of antimicrobial tolerance and resistance in bacterial and fungal biofilms. *Trends Microbiol* **22**, 326-333 (2014). https://doi.org/10.1016/j.tim.2014.02.001
8   O'hara, C., Tenover, F. & Miller, J. Parallel comparison of accuracy of API 20E, Vitek GNI, MicroScan Walk/Away Rapid ID, and Becton Dickinson Cobas Micro ID-E/NF for identification of members of the family Enterobacteriaceae and common gram-negative, non-glucose-fermenting bacilli. *Journal of clinical microbiology* **31**, 3165-3169 (1993).
9   Moll, W. M., Ungerechts, J., Marklein, G. & Schaal, K. P. Comparison of BBL crystal® ANR ID kit and API rapid ID 32 A for identification of anaerobic bacteria. *Zentralblatt für Bakteriologie* **284**, 329-347 (1996).
10  Lamy, B. *et al.* Accuracy of 6 commercial systems for identifying clinical Aeromonas isolates. *Diagnostic microbiology and infectious disease* **67**, 9-14 (2010).
11  Kitch, T. T., Jacobs, M. R. & Appelbaum, P. C. Evaluation of RapID onE system for identification of 379 strains in the family Enterobacteriaceae and oxidase-negative, gram-negative nonfermenters. *Journal of clinical microbiology* **32**, 931-934 (1994).
12  Claxton, n. P. & Masterton, R. Rapid organism identification from Bactec NR blood culture media in a diagnostic microbiology laboratory. *Journal of clinical pathology* **47**, 796-798 (1994).
13  Guarner, J., Street, C., Matlock, M., Cole, L. & Brierre, F. Improving Gram stain proficiency in hospital and satellite laboratories that do not have microbiology. *Clin Chem Lab Med* **55**, 458-461 (2017). https://doi.org/10.1515/cclm-2016-0556
14  Madison, B. M. Application of stains in clinical microbiology. *Biotech Histochem* **76**, 119-125 (2001).
15  Barghouthi, S. A. A universal method for the identification of bacteria based on general PCR primers. *Indian journal of microbiology* **51**, 430-444 (2011).
16  Kuch, A., Stefaniuk, E., Ozorowski, T. & Hryniewicz, W. New selective and differential chromogenic agar medium, chromID VRE, for screening vancomycin-resistant Enterococcus species. *Journal of microbiological methods* **77**, 124-126 (2009).

17    Peterson, J. F. *et al.* Spectra MRSA, a new chromogenic agar medium to screen for methicillin-resistant Staphylococcus aureus. *Journal of clinical microbiology* **48**, 215-219 (2010).
18    Juan, C. *et al.* Challenges for accurate susceptibility testing, detection and interpretation of β-lactam resistance phenotypes in Pseudomonas aeruginosa: results from a Spanish multicentre study. *Journal of Antimicrobial Chemotherapy* **68**, 619-630 (2013).
19    Khan, A. *et al.* Evaluation of the Vitek 2, Phoenix, and MicroScan for antimicrobial susceptibility testing of Stenotrophomonas maltophilia. *Journal of Clinical Microbiology* **59**, e00654-00621 (2021).
20    Choi, Y. *et al.* Scanner-free and wide-field endoscopic imaging by using a single multimode optical fiber. *Phys Rev Lett* **109**, 203901 (2012). https://doi.org/10.1103/PhysRevLett.109.203901
21    Chapin, K. C. & Musgnug, M. C. Validation of the automated reading and incubation system with Sensititre plates for antimicrobial susceptibility testing. *Journal of clinical microbiology* **41**, 1951-1956 (2003).
22    Ferrand, J. *et al.* Comparison of Vitek MS and MALDI Biotyper for identification of Actinomycetaceae of clinical importance. *Journal of Clinical Microbiology* **54**, 782-784 (2016).
23    Carbonnelle, E. *et al.* MALDI-TOF mass spectrometry tools for bacterial identification in clinical microbiology laboratory. *Clinical biochemistry* **44**, 104-109 (2011).
24    Nguyen, M. H. *et al.* Performance of the T2Bacteria panel for diagnosing bloodstream infections: a diagnostic accuracy study. *Annals of internal medicine* **170**, 845-852 (2019).
25    Zboromyrska, Y. *et al.* Evaluation of the Magicplex™ Sepsis real-time test for the rapid diagnosis of bloodstream infections in adults. *Frontiers in cellular and infection microbiology* **9**, 56 (2019).
26    Parize, P. *et al.* Untargeted next-generation sequencing-based first-line diagnosis of infection in immunocompromised adults: a multicentre, blinded, prospective study. *Clinical Microbiology and Infection* **23**, 574. e571-574. e576 (2017).
27    Dark, P. *et al.* Accuracy of LightCycler® Septi F ast for the detection and identification of pathogens in the blood of patients with suspected sepsis: a systematic review and meta-analysis. *Intensive care medicine* **41**, 21-33 (2015).
28    Stevenson, M. *et al.* Sepsis: the LightCycler SeptiFast Test MGRADE®, SepsiTest™ and IRIDICA BAC BSI assay for rapidly identifying bloodstream bacteria and fungi-a systematic review and economic evaluation. *Health Technology Assessment (Winchester, England)* **20**, 1-246 (2016).
29    Blaschke, A. J. *et al.* Rapid identification of pathogens from positive blood cultures by multiplex polymerase chain reaction using the FilmArray system. *Diagnostic microbiology and infectious disease* **74**, 349-355 (2012).
30    Spencer, D. H., Sellenriek, P. & Burnham, C.-A. D. Validation and implementation of the GeneXpert MRSA/SA blood culture assay in a pediatric setting. *American journal of clinical pathology* **136**, 690-694 (2011).
31    Cenci, E. *et al.* Accelerate Pheno™ blood culture detection system: a literature review. *Future Microbiology* **15**, 1595-1605 (2020).
32    Kim, J. H. *et al.* Direct rapid antibiotic susceptibility test (dRAST) for blood culture and its potential usefulness in clinical practice. *J Med Microbiol* **67**, 325-331 (2018). https://doi.org/10.1099/jmm.0.000678
33    Alvarez-Ordonez, A., Mouwen, D., Lopez, M. & Prieto, M. Fourier transform infrared spectroscopy as a tool to characterize molecular composition and stress response in foodborne pathogenic bacteria. *Journal of microbiological methods* **84**, 369-378 (2011).
34    Buzalewicz, I. *et al.* Integrated multi-channel optical system for bacteria characterization and its potential use for monitoring of environmental bacteria. *Biomedical Optics Express* **10**, 1165-1183 (2019).
35    Tasaki, S., Nakayama, M. & Shoji, W. Self-organization of bacterial communities against environmental pH variation: Controlled chemotactic motility arranges cell population structures in biofilms. *PloS one* **12**, e0173195 (2017).

36   Beal, J. *et al.* Robust estimation of bacterial cell count from optical density. *Communications biology* **3**, 1-29 (2020).
37   Pawley, J. *Handbook of biological confocal microscopy*. Vol. 236 (Springer Science & Business Media, 2006).
38   Huisken, J., Swoger, J., Del Bene, F., Wittbrodt, J. & Stelzer, E. H. Optical sectioning deep inside live embryos by selective plane illumination microscopy. *Science* **305**, 1007-1009 (2004).
39   Chen, B.-C. *et al.* Lattice light-sheet microscopy: imaging molecules to embryos at high spatiotemporal resolution. *Science* **346**, 1257998 (2014).
40   Gustafsson, M. G. Nonlinear structured-illumination microscopy: wide-field fluorescence imaging with theoretically unlimited resolution. *Proceedings of the National Academy of Sciences* **102**, 13081-13086 (2005).
41   Li, D. *et al.* Extended-resolution structured illumination imaging of endocytic and cytoskeletal dynamics. *Science* **349**, aab3500 (2015).
42   Hell, S. W. & Wichmann, J. Breaking the diffraction resolution limit by stimulated emission: stimulated-emission-depletion fluorescence microscopy. *Opt. Lett.* **19**, 780-782 (1994).
43   Morozova, K. S. *et al.* Far-red fluorescent protein excitable with red lasers for flow cytometry and superresolution STED nanoscopy. *Biophysical journal* **99**, L13-L15 (2010).
44   Hofmann, M., Eggeling, C., Jakobs, S. & Hell, S. W. Breaking the diffraction barrier in fluorescence microscopy at low light intensities by using reversibly photoswitchable proteins. *Proceedings of the National Academy of Sciences* **102**, 17565-17569 (2005).
45   Betzig, E. *et al.* Imaging intracellular fluorescent proteins at nanometer resolution. *science* **313**, 1642-1645 (2006).
46   Rust, M. J., Bates, M. & Zhuang, X. Sub-diffraction-limit imaging by stochastic optical reconstruction microscopy (STORM). *Nature methods* **3**, 793-796 (2006).
47   Stracy, M. *et al.* Live-cell superresolution microscopy reveals the organization of RNA polymerase in the bacterial nucleoid. *Proceedings of the National Academy of Sciences* **112**, E4390-E4399 (2015).
48   Biteen, J. S. *et al.* Super-resolution imaging in live Caulobacter crescentus cells using photoswitchable EYFP. *Nature methods* **5**, 947-949 (2008).
49   Mir, M. *et al.* Optical measurement of cycle-dependent cell growth. *Proceedings of the National Academy of Sciences* **108**, 13124-13129 (2011).
50   Badrinarayanan, A., Reyes-Lamothe, R., Uphoff, S., Leake, M. C. & Sherratt, D. J. In vivo architecture and action of bacterial structural maintenance of chromosome proteins. *Science* **338**, 528-531 (2012).
51   Vink, J. N. *et al.* Direct visualization of native CRISPR target search in live bacteria reveals cascade DNA surveillance mechanism. *Molecular Cell* **77**, 39-50. e10 (2020).
52   García-Bayona, L. *et al.* Nanaerobic growth enables direct visualization of dynamic cellular processes in human gut symbionts. *Proceedings of the National Academy of Sciences* **117**, 24484-24493 (2020).
53   Zernike, F. How I Discovered Phase Contrast. *Science* **121**, 345-349 (1955). https://doi.org/10.1126/science.121.3141.345
54   Allen, R. & David, G. The Zeiss-Nomarski differential interference equipment for transmitted-light microscopy. *Zeitschrift fur wissenschaftliche Mikroskopie und mikroskopische Technik* **69**, 193-221 (1969).
55   Koenig, K., Hibst, R., Meyer, H., Flemming, G. & Schneckenburger, H. in *Dental applications of lasers.*   170-180 (Spie).
56   Bhartia, R. *et al.* Label-free bacterial imaging with deep-UV-laser-induced native fluorescence. *Applied and environmental microbiology* **76**, 7231-7237 (2010).
57   Datta, R., Heaster, T. M., Sharick, J. T., Gillette, A. A. & Skala, M. C. Fluorescence lifetime imaging microscopy: fundamentals and advances in instrumentation, analysis, and applications. *Journal of biomedical optics* **25**, 071203 (2020).
58   Pereira, F. M. V. *et al.* Laser-induced fluorescence imaging method to monitor citrus greening disease. *Computers and electronics in agriculture* **79**, 90-93 (2011).

59 Hariri, L. P. *et al.* Ex vivo optical coherence tomography and laser-induced fluorescence spectroscopy imaging of murine gastrointestinal tract. *Comparative Medicine* **57**, 175-185 (2007).
60 Zumbusch, A., Holtom, G. R. & Xie, X. S. Three-Dimensional Vibrational Imaging by Coherent Anti-Stokes Raman Scattering. *Physical Review Letters* **82**, 4142-4145 (1999). https://doi.org/10.1103/PhysRevLett.82.4142
61 Freudiger, C. W. *et al.* Label-Free Biomedical Imaging with High Sensitivity by Stimulated Raman Scattering Microscopy. *Science* **322**, 1857 (2008). https://doi.org/10.1126/science.1165758
62 Hong, W. *et al.* Antibiotic susceptibility determination within one cell cycle at single-bacterium level by stimulated Raman metabolic imaging. *Analytical chemistry* **90**, 3737-3743 (2018).
63 Zhang, M. *et al.* Rapid determination of antimicrobial susceptibility by stimulated Raman scattering imaging of D2O metabolic incorporation in a single bacterium. *Advanced Science* **7**, 2001452 (2020).
64 Zhang, W. *et al.* Rapid antimicrobial susceptibility testing by stimulated Raman scattering metabolic imaging and morphological deformation of bacteria. *Analytica Chimica Acta* **1168**, 338622 (2021).
65 Li, J. *et al.* Three-dimensional tomographic microscopy technique with multi-frequency combination with partially coherent illuminations. *Biomedical Optics Express* **9**, 2526-2542 (2018). https://doi.org/10.1364/BOE.9.002526
66 Kang, I., Pang, S., Zhang, Q., Fang, N. & Barbastathis, G. Recurrent neural network reveals transparent objects through scattering media. *Optics Express* **29**, 5316-5326 (2021). https://doi.org/10.1364/OE.412890
67 Jarvis, R. M., Brooker, A. & Goodacre, R. Surface-enhanced Raman spectroscopy for bacterial discrimination utilizing a scanning electron microscope with a Raman spectroscopy interface. *Analytical Chemistry* **76**, 5198-5202 (2004).
68 Park, Y., Depeursinge, C. & Popescu, G. Quantitative phase imaging in biomedicine. *Nature photonics* **12**, 578-589 (2018).
69 Kim, G. *et al.* Holotomography. *Nature Reviews Methods Primers* **4**, 51 (2024).
70 Jo, Y. *et al.* Angle-resolved light scattering of individual rod-shaped bacteria based on Fourier transform light scattering. *Scientific reports* **4**, 1-6 (2014).
71 Shin, J., Kim, G., Park, J., Lee, M. & Park, Y. Long-term label-free assessments of individual bacteria using three-dimensional quantitative phase imaging and hydrogel-based immobilization. *Scientific Reports* **13**, 46 (2023).
72 Choi, S. Y., Oh, J., Jung, J., Park, Y. & Lee, S. Y. Three-dimensional label-free visualization and quantification of polyhydroxyalkanoates in individual bacterial cell in its native state. *Proceedings of the National Academy of Sciences* **118**, e2103956118 (2021).
73 Kim, T. I. *et al.* Antibacterial Activities of Graphene Oxide–Molybdenum Disulfide Nanocomposite Films. *ACS Applied Materials & Interfaces* **9**, 7908-7917 (2017). https://doi.org/10.1021/acsami.6b12464
74 Oh, J. *et al.* Three-dimensional label-free observation of individual bacteria upon antibiotic treatment using optical diffraction tomography. *Biomedical optics express* **11**, 1257-1267 (2020).
75 Jo, Y. *et al.* Holographic deep learning for rapid optical screening of anthrax spores. *Science advances* **3**, e1700606 (2017).
76 Kim, G. *et al.* Rapid species identification of pathogenic bacteria from a minute quantity exploiting three-dimensional quantitative phase imaging and artificial neural network. *Light: Science & Applications* **11**, 190 (2022).
77 Moriarty-Craige, S. E. & Jones, D. P. Extracellular thiols and thiol/disulfide redox in metabolism. *Annu. Rev. Nutr.* **24**, 481-509 (2004).
78 Srinivasan, B. *et al.* TEER measurement techniques for in vitro barrier model systems. *Journal of laboratory automation* **20**, 107-126 (2015).
79 Lee, R. *et al.* Dielectric imaging for differentiation between cancer and inflammation in vivo. *Scientific reports* **7**, 13137 (2017).
19

80  Spencer, D. C. *et al.* A fast impedance-based antimicrobial susceptibility test. *Nature communications* **11**, 5328 (2020).
81  Abbott, J. *et al.* Multi-parametric functional imaging of cell cultures and tissues with a CMOS microelectrode array. *Lab on a Chip* **22**, 1286-1296 (2022).
82  Chitale, S. *et al.* A semiconductor 96-microplate platform for electrical-imaging based high-throughput phenotypic screening. *Nature Communications* **14**, 7576 (2023).
83  Incandela, J. T., Hu, K., Joshi, P., Rosenstein, J. K. & Larkin, J. W. Non-Optical, Label-free Electrical Capacitance Imaging of Microorganisms. *bioRxiv*, 2024.2009. 2025.615041 (2024).
84  Gey, A., Werckenthin, C., Poppert, S. & Straubinger, R. K. Identification of pathogens in mastitis milk samples with fluorescent in situ hybridization. *J Vet Diagn Invest* **25**, 386-394 (2013). https://doi.org/10.1177/1040638713486113
85  Bhattacharjee, A., Datta, R., Gratton, E. & Hochbaum, A. I. Metabolic fingerprinting of bacteria by fluorescence lifetime imaging microscopy. *Scientific reports* **7**, 1-10 (2017).
86  Singhal, N., Kumar, M., Kanaujia, P. K. & Virdi, J. S. MALDI-TOF mass spectrometry: an emerging technology for microbial identification and diagnosis. *Frontiers in microbiology* **6**, 791 (2015).
87  Amann, R. & Fuchs, B. M. Single-cell identification in microbial communities by improved fluorescence in situ hybridization techniques. *Nature Reviews Microbiology* **6**, 339-348 (2008).
88  Frickmann, H. *et al.* Fluorescence in situ hybridization (FISH) in the microbiological diagnostic routine laboratory: a review. *Critical reviews in microbiology* **43**, 263-293 (2017).
89  Müller, V. *et al.* Identification of pathogenic bacteria in complex samples using a smartphone based fluorescence microscope. *RSC advances* **8**, 36493-36502 (2018).
90  Patino, S. *et al.* Autofluorescence of mycobacteria as a tool for detection of Mycobacterium tuberculosis. *Journal of Clinical Microbiology* **46**, 3296-3302 (2008).
91  Jo, Y. *et al.* Label-free identification of individual bacteria using Fourier transform light scattering. *Optics express* **23**, 15792-15805 (2015).
92  Son, K., Guasto, J. S. & Stocker, R. Bacteria can exploit a flagellar buckling instability to change direction. *Nature physics* **9**, 494-498 (2013).
93  Hillman, T. R. *et al.* Digital optical phase conjugation for delivering two-dimensional images through turbid media. *Scientific Reports* **3**, 1-5 (2013). https://doi.org/10.1038/srep01909
94  Taute, K., Gude, S., Tans, S. & Shimizu, T. High-throughput 3D tracking of bacteria on a standard phase contrast microscope. *Nature communications* **6**, 1-9 (2015).
95  Son, K., Brumley, D. R. & Stocker, R. Live from under the lens: exploring microbial motility with dynamic imaging and microfluidics. *Nature Reviews Microbiology* **13**, 761-775 (2015).
96  Vissers, T. *et al.* Dynamical analysis of bacteria in microscopy movies. *Plos one* **14**, e0217823 (2019).
97  Acres, J. & Nadeau, J. 2D vs 3D tracking in bacterial motility analysis. *Biophysics* (2021).
98  Hyon, Y., Powers, T. R., Stocker, R. & Fu, H. C. The wiggling trajectories of bacteria. *Journal of Fluid Mechanics* **705**, 58-76 (2012).
99  Lee, K. C. *et al.* Quantitative phase imaging flow cytometry for ultra-large-scale single-cell biophysical phenotyping. *Cytometry Part A* **95**, 510-520 (2019).
100 Kamdar, S. *et al.* The colloidal nature of complex fluids enhances bacterial motility. *Nature* **603**, 819-823 (2022).
101 Makarchuk, S., Braz, V. C., Araújo, N. A., Ciric, L. & Volpe, G. Enhanced propagation of motile bacteria on surfaces due to forward scattering. *Nature communications* **10**, 1-12 (2019).
102 Ariel, G. *et al.* Swarming bacteria migrate by Lévy Walk. *Nature communications* **6**, 1-6 (2015).
103 Cheong, F. C. *et al.* Rapid, high-throughput tracking of bacterial motility in 3D via phase-contrast holographic video microscopy. *Biophysical journal* **108**, 1248-1256 (2015).
104 Molaei, M. & Sheng, J. Imaging bacterial 3D motion using digital in-line holographic microscopy and correlation-based de-noising algorithm. *Optics express* **22**, 32119-32137 (2014).
105 Jericho, S. *et al.* In-line digital holographic microscopy for terrestrial and exobiological research. *Planetary and Space Science* **58**, 701-705 (2010).

106	Qin, B. *et al.* Cell position fates and collective fountain flow in bacterial biofilms revealed by light-sheet microscopy. *Science* **369**, 71-77 (2020).
107	Zhang, M. *et al.* Non-invasive single-cell morphometry in living bacterial biofilms. *Nature communications* **11**, 1-13 (2020).
108	Manzo, N. *et al.* Pigmentation and sporulation are alternative cell fates in Bacillus pumilus SF214. *PLoS One* **8**, e62093 (2013).
109	Wu, F. *et al.* Direct imaging of the circular chromosome in a live bacterium. *Nature communications* **10**, 1-9 (2019).
110	Dewasthale, S., Mani, I. & Vasdev, K. Microbial biofilm: current challenges in health care industry. *J Appl Biotechnol Bioeng* **5**, 160-164 (2018).
111	Hartmann, R. *et al.* Quantitative image analysis of microbial communities with BiofilmQ. *Nature microbiology* **6**, 151-156 (2021).
112	Basu, R. *et al.* Cytotoxic T Cells Use Mechanical Force to Potentiate Target Cell Killing. *Cell* **165**, 100-110 (2016). https://doi.org/10.1016/j.cell.2016.01.021
113	Díaz-Pascual, F. *et al.* Breakdown of Vibrio cholerae biofilm architecture induced by antibiotics disrupts community barrier function. *Nature microbiology* **4**, 2136-2145 (2019).
114	Giri, S., Waschina, S., Kaleta, C. & Kost, C. Defining division of labor in microbial communities. *Journal of molecular biology* **431**, 4712-4731 (2019).
115	Momeni, B. Division of labor: how microbes split their responsibility. *Current Biology* **28**, R697-R699 (2018).
116	Moradali, M. F. & Rehm, B. H. A. Bacterial biopolymers: from pathogenesis to advanced materials. *Nature Reviews Microbiology* **18**, 195-210 (2020). https://doi.org/10.1038/s41579-019-0313-3
117	Choi, S. Y., Oh, J., Jung, J., Park, Y. & Lee, S. Y. Three-dimensional label-free visualization and quantification of polyhydroxyalkanoates in individual bacterial cell in its native state. *Proceedings of the National Academy of Sciences* **118** (2021).
118	Lu, Y. *et al.* Single cell antimicrobial susceptibility testing by confined microchannels and electrokinetic loading. *Analytical chemistry* **85**, 3971-3976 (2013).
119	Mohan, R. *et al.* A multiplexed microfluidic platform for rapid antibiotic susceptibility testing. *Biosensors and Bioelectronics* **49**, 118-125 (2013).
120	Baltekin, Ö., Boucharin, A., Tano, E., Andersson, D. I. & Elf, J. Antibiotic susceptibility testing in less than 30 min using direct single-cell imaging. *Proceedings of the National Academy of Sciences* **114**, 9170-9175 (2017).
121	Jung, Y.-G. *et al.* A rapid culture system uninfluenced by an inoculum effect increases reliability and convenience for drug susceptibility testing of Mycobacterium tuberculosis. *Scientific reports* **8**, 1-11 (2018).
122	Ban, S. *et al.* Optical properties of acute kidney injury measured by quantitative phase imaging. *Biomedical optics express* **9**, 921-932 (2018).
123	Matsumoto, Y. *et al.* A microfluidic channel method for rapid drug-susceptibility testing of Pseudomonas aeruginosa. *PloS one* **11**, e0148797 (2016).
124	Veses-Garcia, M. *et al.* Rapid phenotypic antibiotic susceptibility testing of uropathogens using optical signal analysis on the nanowell slide. *Frontiers in microbiology* **9**, 1530 (2018).
125	Jung, Y. G. *et al.* A rapid culture system uninfluenced by an inoculum effect increases reliability and convenience for drug susceptibility testing of Mycobacterium tuberculosis. *Sci Rep* **8**, 8651 (2018). https://doi.org/10.1038/s41598-018-26419-z
126	Kim, T. H. *et al.* Blood culture-free ultra-rapid antimicrobial susceptibility testing. *Nature* **632**, 893-902 (2024).
127	Dare, R. K. *et al.* Clinical impact of Accelerate Pheno rapid blood culture detection system in bacteremic patients. *Clinical Infectious Diseases* **73**, e4616-e4626 (2021).
128	Descours, G. *et al.* Evaluation of the Accelerate Pheno™ system for rapid identification and antimicrobial susceptibility testing of Gram-negative bacteria in bloodstream infections. *European Journal of Clinical Microbiology & Infectious Diseases* **37**, 1573-1583 (2018).
129	Zhang, K., Qin, S., Wu, S., Liang, Y. & Li, J. Microfluidic systems for rapid antibiotic susceptibility tests (ASTs) at the single-cell level. *Chemical science* **11**, 6352-6361 (2020).

| | |
|---|---|
| 130 | Kim, S., Lee, S., Kim, J.-K., Chung, H. J. & Jeon, J. S. Microfluidic-based observation of local bacterial density under antimicrobial concentration gradient for rapid antibiotic susceptibility testing. *Biomicrofluidics* **13** (2019). |
| 131 | Kim, K., Kim, S. & Jeon, J. S. Visual estimation of bacterial growth level in microfluidic culture systems. *Sensors* **18**, 447 (2018). |
| 132 | He, K., Zhang, X., Ren, S. & Sun, J. in *Proceedings of the IEEE conference on computer vision and pattern recognition.*   770-778. |
| 133 | Huang, G., Liu, Z., Van Der Maaten, L. & Weinberger, K. Q. in *Proceedings of the IEEE conference on computer vision and pattern recognition.*   4700-4708. |
| 134 | Zhou, B., Khosla, A., Lapedriza, A., Oliva, A. & Torralba, A. in *Proceedings of the IEEE conference on computer vision and pattern recognition.*   2921-2929. |
| 135 | Selvaraju, R. R. *et al.* in *Proceedings of the IEEE international conference on computer vision.* 618-626. |
| 136 | Kendall, A. & Gal, Y. What uncertainties do we need in bayesian deep learning for computer vision? *Advances in neural information processing systems* **30** (2017). |
| 137 | Spahn, C. *et al.* DeepBacs for multi-task bacterial image analysis using open-source deep learning approaches. *Communications biology* **5**, 1-18 (2022). |